\documentstyle[osa,manuscript,graphicx]{revtex}
\begin{document}
%\setcounter{page}{38}

%%%%%%%%%%%%%%%%%%%%%%%%%%%%%%%%
\begin{center}{\Large \bf  Effects of Extra Space-time Dimensions 
on the Fermi Constant } \\
\vskip.25in
{Pran Nath$^a$ and Masahiro Yamaguchi$^b$ }

{\it
a Department of Physics, Northeastern University, Boston, MA 02115, USA\\
b Department of Physics, Tohoku University, Sendai 980-8578, Japan
} \\%

\end{center}

\begin{abstract}                % DON'T CHANGE THIS LINE
Effects of Kaluza-Klein excitations associated with 
extra dimensions with large radius compactifications
 on the Fermi constant are explored. 
 It is  shown that  the current precision determinations of the
 Fermi constant, of the fine structure constant, and 
  of the W and Z mass put stringent 
 constraints on the compactification radius. The analysis  
 excludes 
 one extra space time dimension below $\sim 1.6$ TeV, and 
 excludes 2, 3 and 4 extra
 space dimensions opening simultaneously below $\sim$ 
 3.5 TeV, 5.7 TeV and 7.8 TeV at the $90\% ~CL$.
Implications of these results for future collider  
 experiments are discussed. 
 \end{abstract}
The Kaluza-Klein theories have a long and rich 
history\cite{kaluza,freund}. 
 The TeV scale
 strings provide a new impetus for studying these theories in the 
 context of low energy phenomenology.
Till recently  much of the string phenomenology has been 
conducted in the framework of the weakly coupled heterotic string
where a rigid relationship exists between the string scale ($M_{str}$)
and the Planck scale $M_{Planck}$\cite{kapu}, i.e., 
$M_{str}\sim g_{str} M_{Planck}$
where $M_{Planck}=(8\pi G_N)^{-1/2}$, 
and where $G_N$ is the Newton's constant and numerically
$M_{Planck}=1.2\times 10^{19}$ GeV. However, recently the advent of 
string dualities has opened the new possibility which relates the 
strong coupling regime of one string theory to the weak coupling 
limit of another. Thus it is conjectured that the strongly coupled 
SO(32) heterotic string compactified to four dimensions is equivalent to 
a weakly coupled Type I string compactified on four dimensions. 
 In this context the   string scale  can be 
very different\cite{witten,lykken}.
While one is very far from generating 
 realistic TeV string models the generic features of models of this type
with low values of the string scales can nonetheless be 
studied\cite{anto,dienes,kaku}.
Here we adopt the picture that matter resides in D=d+4 dimensions
while gravity propagates in the 10 dimensional bulk. In the context
of Type I strings one may conjecture that matter resides on a p-brane 
(p=d+3) 
with compactification of d dimensions occuring internal to the brane
while the compactification of the remaining (6-d) dimensions occurs 
in directions transverse to the p-brane. 
We focus on the 
d compactifications internal to the brane and the compactifications 
transverse to the brane will not concern us here. We shall work
within the framework of an effective field theory, which is a valid 
approximation in the domain of energy investigated below.
 The main motivation of this Letter is to 
analyse the effects of the Kaluza-Klein excitations associated with
the extra dimensions on the Fermi constant $G_F$ which is one of
the most accurately determined quantities in particle 
physics\cite{ritbergen}. 
In our analysis 
we shall assume that a number d of extra dimensions open at 
a low scale each associated 
with a common radius of compactification R=$M_R^{-1}$.

We consider the 5D  case first. 
 Our starting point is the $MSSM$ Lagrangian in 5D
 and we exhibit here a few terms to define notation
 \begin{equation}
L_5=-\frac{1}{4} F_{MN}F^{MN}
-(D_MH_i)^{\dagger}(D^MH_i)
-\bar\psi\frac{1}{i}\Gamma^M D_M\psi-V(H_i)+..
\end{equation}
where M,N are the five dimensional indices that run over values 
0,1,2,3,5, $H_i$ (i=1,2) stand for two Higgs hypermultiplets which
contain the MSSM Higgs,
$D_M=\partial_M-ig^{(5)}A_M$, where 
 $g^{(5)}$ are the gauge 
coupling constants and the $SU(3)\times SU(2)\times U(1)$ indices are
suppressed, and $ V(H_i)$ is the Higgs potential. 
  This theory is constructed to have  N=1
 supersymmetry in 5D. We compactify this theory on $S^1/Z_2$ with 
 the radius of compatification R. 
 In our analysis we assume that 
 the gauge fields and the Higgs fields 
 live in the five dimensional bulk while the quarks and leptons are 
 confined to a 4 dimensional wall, i.e, at a $Z_2$  
 fixed point\cite{couplings}. 
After compactification 
  the resulting
 spectrum contains massless modes with N=1 supersymmetry in 4D, which 
  precisely form the spectrum of MSSM in 4D,  and   
 massive Kaluza-Klein modes which form N=2 multiplets in 4D.
 The modes
in $4D$ can be further labelled as even or odd under the action of $Z_2$. 
The MSSM spectrum and their Kaluza-Klein excitations are even under 
$Z_2$ while the remaining N=1 massive set of Kaluza-Klein fields are
odd. We carry out a spontaneous breaking of the electro-weak 
symmetry in 5D  which gives electro-weak masses to the 
W and Z gauge bosons and to their Kaluza-Klein modes in addition 
to the compatification masses, i.e., Kaluza-Klein modes have 
masses $(m_i^2+n^2 M_R^2)$, n=1,2,3,.., $\infty$, where $m_i^2$ 
arise from electro-weak symmetry breaking and $n^2R^2$ arise
from compactification of the fifth dimension.  
In $4D$ a rescaling of coupling constants is needed, i.e., 
$g^{(5)}_i/{\sqrt{\pi R}}=g_i$
where $g_i$ are the gauge coupling constants in four dimensions.
The interactions of the 
fermions to zero modes and to Kaluza-Klein modes after rescaling
is given by
\begin{equation}
{\it L_{int}}=g_ij^{\mu}(A_{\mu i}+\sqrt 2\sum_{n=1}^{\infty} A_{\mu i}^n)
\end{equation}
where $A_{\mu i}$ are the zero modes and  $A_{\mu i}^n$ are the 
Kaluza-Klein modes. 

The Fermi constant is very accurately known from the weak interaction 
processes. Its current experimental value obtained from the muon
lifetime including the complete 2-loop quantum electromagnetic 
contributions is\cite{ritbergen}
\begin{equation}
G_F=1.16639(1)\times 10^{-5} ~GeV^{-2}
\end{equation}
where the number in the parenthesis is the error.
We compare now the experimental result on $G_F$ of Eq.(3) with the
predictions of $G_F$ of Eq.(3) in the Standard Model.
 In the on-shell scheme $G_F^{SM}$ is given by\cite{sirlin} 
\begin{equation}
G_F^{SM}=\frac{\pi\alpha}{\sqrt 2M_W^2 sin^2\theta_W (1-\Delta r)}
 \end{equation}
 where $sin^2\theta_W=(1-M_W^2/M_Z^2)$ in the on-shell scheme. 
 The fine structure constant $\alpha$ (at $Q^2=0$) is very accurately
 known, i.e., $\alpha^{-1}=137.0359$, and  $\Delta r$ is the radiative
 correction and is determined to be\cite{pdg} $\Delta r=0.0349\mp 0.0019
 \pm 0.0007$ where the first error comes from the error in the mass of
 the top quark ($m_t=175\pm 5$ GeV) and the second error comes from 
 the uncertainty of
 $\alpha (M_Z)$. With the above one can use the measured values of
 $M_W$ and $M_Z$ to derive the value of $G_F^{(SM)}$. The currently 
 measured values of $M_W$ and $M_Z$ are 
 $M_W=80.39\pm 0.06$ GeV\cite{karlin} and
 $M_Z=91.1867\pm 0.0020$ GeV\cite{pdg}. 
 Using the above determinations one finds
 that the Standard Model prediction of $G_F$ is in excellent accord 
 with experiment. Thus we 
 conclude that the contribution of the Kaluza-Klein modes must fall 
 in the error corridors of the Standard Model to be consistent with
 current experiment. Since the fine structure constant and $M_Z$
 are very accurately known most of the error in the evaluation of
 $G_F^{SM}$ arises from the errors in the measurement of the W mass
 and in the evaluation of the radiative correction $\Delta r$. 
 To exhibit the relative contribution of the errors from these 
 sources to $G_F$ we find that  
\begin{equation}
\Delta G_F/G_F|_{SM}\simeq \sqrt{4(1/sin^2\theta_W-2)^2 
(\delta M_W/M_W)^2
+(\delta \Delta r)^2}
\end{equation} 
The quantity $2(1/sin^2\theta_W-2)\sim 5$ is 
 accidentally large and thus the error in $M_W$ dominates the
 error in the Standard Model contribution to the Fermi constant.
   Using the above analysis we find that 
  $G_F^{SM}$
 =$(1.16775\pm 0.0049)\times 10^{-5}$ $GeV^{-2}$.
We note that while the error in the measurement of the W mass is 
less than 0.1\% it gets enlarged to around 0.5\% due to the 
enhancement factor of 5.
Even so the error corridor of $G_F^{SM}$ is very narrow and 
places a strong constraint on new physics.
 Thus we use this corridor to constrain the 
 Kaluza-Klein contributions to the Fermi constant.
 In our analysis we shall assume that the radiative corrections
 from the Kaluza-Klein states to $\Delta r$ are small. This 
 assumption will turn out to be {\it aposteriori} justified in view
 of the largeness of the limits on masses of the Kaluza-Klein 
 excitations obtained in this analysis. Under the above assumption
   we then require that  $G_F^{KK}$ be limited by the error in 
  $\Delta G_F^{SM}$, i.e.,
\begin{equation}
\Delta G_F^{KK}/G_F^{SM}\leq \pm 0.82\times 10^{-2}~~~~~(90\%~CL)
\end{equation} 

For the case of one extra dimension, 
after compactification and integration over the W boson and its
Kaluza-Klein excitations we obtain the effective Fermi constant  
to leading order in $M_W/M_R$ to be  
\begin{equation}
G_F^{eff}\simeq G_F^{SM}(1+\frac{\pi^2}{3}\frac{M_W^2}{M_R^2})
\end{equation}
Defining $\Delta G_F^{KK}=G_F^{eff}-G_F^{SM}$, identifying $G_F^{eff}$
 with the experimental value of the Fermi constant,  
and using Eq.(6) we find $M_R >1.6$ TeV ($90\%~CL$).  
This limit on $M_R$ is stronger than for the case
when one has just an extra W recurrence. Thus if one had an extra
W recurrence with a mass $M_{W'}$, the analysis above will give a limit
$M_{W'}>905$ GeV. 
The stronger limit for the Kaluza-Klein case
is due to an enhancement factor of ${\pi}/\sqrt {3}$. This factor 
arises in part due to 
summation over the tower of Kaluza-Klein states and in part due 
 the coupling of the 4D fermions to the 
Kaluza-Klein gauge bosons being stronger by a factor of $\sqrt 2$
relative to the couplings of the fermions to the zero mode gauge 
bosons as can be seen from Eq.(2).
The current experimental limit on the recurrance of
a W is $M_{W'}>720$ GeV given by the analysis at 
 the Fermilab Tevatron $p\bar p$ collider data with the 
 D0 detector\cite{abachi}.Thus our result on $M_R$ is  much
 stronger than the current experimental limit on $M_{W'}$.

Another independent constraint on $M_R$  can
be gotten from atomic parity experiments. The atomic parity violations
arise from the Z exchange and the low energy effective interaction
governing the violation is\cite{lang} $L_{PV}^{SM}=(G_F^{SM}/\sqrt 2)$
$\sum_iC_{1i}\bar e\gamma_{\mu}\gamma_5e\bar q_i\gamma^{\mu}q_i$.  
In SM the measured quantity is $Q_W^{SM}$=$2[(2Z+N)C_{1u}$+$
(Z+2N)C_{1d}]$ where Z is the number of protons and N is the number 
of neutron in the atomic nucleus being considered. The exchange of
the Kaluza-Klein Z bosons contribute an additional interaction, i.e., 
\begin{equation} 
 L_{PV}^{KK}=(\Delta G_F^{KK}/\sqrt 2)
\sum_iC_{1i}\bar e\gamma_{\mu}\gamma_5e\bar q_i\gamma^{\mu}q_i
\end{equation}
The most accurate atomic parity experiment is for cesium which gives

\begin{equation}
Q_W^{exp}(Cs)=-72.41\pm 0.25\pm 0.80
\end{equation}
while the Standard Model gives 
\begin{equation}
Q_W^{SM}(Cs)=-73.12\pm 0.06
\end{equation}
The above result gives $\Delta Q_W^{exp}-Q_M^{SM}$=$0.71\pm 0.84$ 
where we have added the errors in quadrature. We define the
Kaluza-Klein contribution to $Q_W$ by $\Delta Q_W^{KK}$=
$(\Delta G_F^{KK}/G_F^{SM})$$Q_W^{(SM)}$ and require that 
$\Delta Q_W^{KK}$ be limited by $\Delta Q_W$, i.e.,
$\Delta Q_W^{KK}\leq \Delta Q_W$. This leads to  
 a constraint on $M_R$ of $M_R> 1.4$ TeV (90\% CL) which 
is less strong than the limit gotten from the analysis of $G_F$ 
but still quite impressive.  The current accuracy of the experimental 
determinations of other electro-weak quantities produce less 
stringent constraints on $M_R$. 

%\section{Extension to d Extra Dimensions}
The above analysis can be extended to d extra dimensions.
We consider here only compacifications with common compactification 
radius R and $Z_2$ type orbifolding.
Integration on the W boson and its Kaluza-Klein excitations 
similar to the 5D case gives the following result for
 the sum of the Standard Model and Kaluza-Klein mode contributions 
\begin{equation}
G_F^{eff}=G_F^{SM}K_d(\frac{M_W^2}{M_R^2})
\end{equation}
where $K_d$ is the Kaluza-Klein dressing of the Fermi constant 
\begin{equation}
K_d(c)=1+\sum_{p=1}^{d}(2^p~^dC_p) C_p(c)
\end{equation}
Here $^dC_p=d!/p!(d-p)!$,  $C_p(c)=\sum_{\vec k_p}(c/(c+\vec k_p^2))$
where $\vec k_p=(k_1,k_2,..,k_p)$, and $k_i$ run over the positive
integers 1,2, ..$\infty$. We can express 
$K_d(c)$ in terms of the Jacobi function 
\begin{equation}
K_d(c)= \int_{0}^{\infty}dt e^{-t}  
(\theta_3(\frac{it}{c\pi}))^{d}
\end{equation}
where $\theta_3(z)$ for complex z ($Imz>0$) is defined by $\theta_3(z)$=
$\sum_{k=-\infty}^{\infty}$$exp(i\pi k^2 z)$. 
For the case of more than one extra dimension both  the sum of
Eq.(12) and the integral of Eq.(13) diverge. To obtain a convergent 
result the lower limit on the integral must be replaced by a cutoff.
The physical origin of the cutoff is a truncation of the 
sum over the Kaluza-Klein states when their masses exceed the string
scale. With the cutoff we obtain the following approximate expressions
for $\Delta G_F^{KK}/G_F^{SM}$

\begin{equation}
\Delta G_F^{KK}/G_F^{SM}\simeq (\frac{2\pi^2}{3}
+2\pi ln(\frac{M_{str}}{M_R}))(\frac{M_W}{M_R})^2, ~~~~d=2 
\end{equation}

\begin{equation}
\Delta G_F^{KK}/G_F^{SM}\simeq (\frac{d}{d-2})
\frac{\pi^{d/2}}{\Gamma (1+\frac{d}{2})} (\frac{M_{str}}{M_R})^{d-2}
(\frac{M_W}{M_R})^2, ~~~d\geq 3
\end{equation}
Numerically we find that the approximation of Eq.(15) agrees with the
exact result of Eq.(12) to 2-3\% while the approximation of Eq.(14)
is good only to about $10\%$ because of the slow convergence of
the log function in this case.
The quantities $M_R$ and $M_{str}$ are not completely arbitrary but are
constrained by the unification of the gauge couplings.  
 Thus in a TeV scale unification 
the gauge coupling evolution is given by\cite{dienes,kaku}
\begin{equation}
\alpha_i(M_Z)^{-1}=\alpha_U^{-1}+\frac{b_i}{2\pi}ln(\frac{M_R}{M_Z})
-\frac{b_i^{KK}}{2\pi}ln(\frac{M_{str}}{M_R})+ 
\Delta_i
%+\frac{b_i^{KK}}{4\pi d}\frac{\pi^{d/2}}{\Gamma (1+\frac{d}{2})}
%[(\frac{M_{str}}{M_R})^d-1]
\end{equation}
Here $\alpha_U$ is the effective  GUT coupling constant,  
$b_i$=$(-3,1,33/5)$
for $SU(3)_C$$\times$$SU(2)_L$$\times$$U(1)_Y$
 describe the evolution of the gauge couplings from the scale Q to the
scale $M_R$, $b_i^{KK}=(-6,-3,3/5)$ are the $b_i$ minus the contribution 
from the  fermion sector which has no Kaluza-Klein excitations, and 
$\Delta_i$ are
 the corrections arising from the Kaluza-Klein states. 
  The requirement that $\alpha_i(M_Z)$ be compatible with the LEP data
  leads to  constraints on $M_R$ and  
${M_{str}}$. For a given d and $M_R$ the unification of 
$\alpha_1$ and $\alpha_2$ thus fixes the  ratio $\frac{M_{str}}{M_R}$.
In ref.\cite{kaku} an extension to include two additional multiplets
$F_+$ and $F_-$ with $SU(3)_C\times SU(2)_L\times U(1)_Y$ quantum 
numbers $(1,1,+2)$ for $F_+$ and $(1,1,-2)$ for $F_-$ was 
proposed with a mass scale $M_F$ in the range
$M_{EW}\leq M_F\leq M_R$. Inclusion of this multiplet improves the
agreement of $\alpha_3$ with experiment but otherwise 
 does not affect our analysis in any substantial way. 

The result of our analysis on the lower limit on extra dimensions 
under the constraints of unification
of gauge couplings and under the constraint of Eq.(6) is given 
in Fig.1.  For the case of one extra dimension the analysis of Fig.1
shows that the constraint of Eq.(6) puts a lower limit on $M_R$ of
1.6 TeV. This is the same result that we got previously from Eqs.(6)
and (7). This is because in this case the unification of the gauge 
coupling constant constraint gives a  high value of 
$M_{str}/M_R$ so the finite sum on the Kaluza-Klein states 
simulates  to a very good approximation the
sum on the  infinite tower of Kaluza-Klein states. 
 For the case of 2, 3 and 4 
extra dimensions the analysis of Fig.(1) shows that Eq.(6) produces
a lower limit on $M_R$ of 3.5 TeV, 5.7 TeV, and 7.8 TeV.
In these cases the truncation of the Kaluza-Klein tower is rather 
severe as one goes to higher values of d, e.g, 
$M_{str}/M_R \simeq 7.6$ for d=2 and this ratio becomes smaller for larger 
values of d (and further the ratio $M_{str}/M_R$ must be 
progressively  fine tuned more severely as one goes to higher values
of d).  The specific nature of the cutoff imposed introduces an 
uncertainty in the accuracy of the predictions which is $O(1\%)$ for 
d=1 and gets larger for larger values of d up to $O(10\%)$
for d=4. The largeness of the uncertainty for larger values of d 
arises due to the progressively smaller number of Kaluza-Klein states
that are retained in the truncation procedure under 
the unification of the gauge coupling constants constraint. 
 As pointed out earlier the analysis is very sensitive 
to the error in the measurement of the W boson mass. Thus if this 
error on $M_W$ were to decrease by a factor of 2, which is not
an unlikely possibility, then the limit on $M_R$ for d=1 will 
increase to 2.1 TeV, and the limits for d=2, d=3 and d=4 will increase
to 4.4 TeV, 7.1 TeV and 9.5 TeV as can be seen by the intercepts of the
dotted line in Fig.1. 
These results may be contrasted with the previous limits of $\sim 300$ GeV
from an analysis of contact interactions\cite{dienes}.
 Thus none of the Kaluza-Klein excitations of the 
 W or Z boson will become visible at the Tevatron. However, Kaluza-Klein
 excitations may still be accessible at the Large Hadron 
 Colldier (LHC)\cite{abq}. 
 
One may ask to what extent our limits are dependent on the assumption of 
an underlying supersymmetry. Since loop corrections to $\Delta r$ from
the Kaluza-Klein states are ignored, which turns out to be a 
reasonable assumption in view of the largeness of the masses of the
Kaluza-Klein states, the essential constraint from supersymmetry
arises via the cutoff. For the 5D case this constraint is rather mild
but does get severe as one goes to larger number of extra 
dimensions. Thus our conclusion is that the 5D limit is essentially
model independent but the limits deduced for the
 6D-8D cases do require the underlying 
supersymmetric framework and the results here are thus more model dependent. 

Finally we note that the limits on extra dimensions are 
constrained severely by the errors in the Standard Model predictions 
of $G_F$ which are 
dominated by the error in the W mass measurement and sub-dominated
by the error in the top mass measurement. Thus if future experiments
 further  reduce the errors in these measurements,
then one will obtain more stringent bounds on the radius of 
compactification, or if one sees a deviation from the Standard Model
prediction it might be a smoking gun signal for a TeV scale 
compactification. 

This research was supported in part by the National Science Foundation
grant no. PHY-9602074. We thank Professors C.W. Kim and J.E. Kim
for the kind hospitality accorded them for the period of the
visit at the Korean Institute of Advanced Study, Seoul
 where this work was initiated.

\begin{figure}
\begin{center}
\includegraphics[angle=0,width=5.5in]{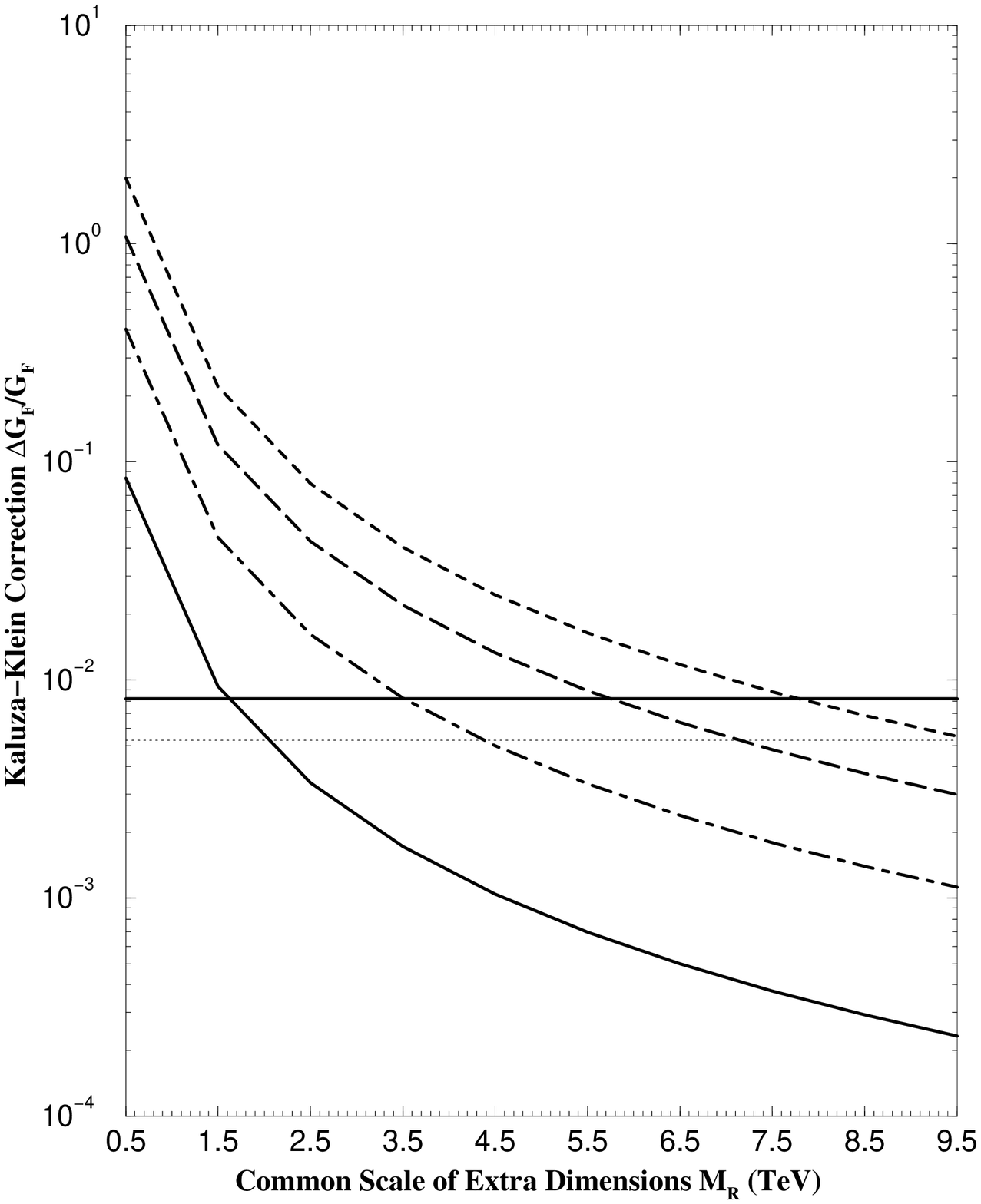}
\caption{Plot of the ratio ($\Delta G_F^{KK}/G_F^{SM}$) as a function
of the common scale of extra dimension. The curves correspond to the
case d=1 (solid), d=2 (dot-dashed), d=3 (long dashed), and d=4
(dashed). The horizontal solid line corresponds to the 2$\sigma$ 
error on  $\Delta G_F^{SM}/G_F^{SM}$  while the horizontal 
dotted line is what the   2$\sigma$  error will be if the error
in the W boson measurements decreased by a factor of 2.}
\label{fig}
\end{center}
\end{figure}


\begin{references}
\bibitem{couplings}
We note that the assumption that the quarks and leptons
 live on the boundary is necessary for them to have non-vanishing
 couplings to the Kaluza-Klein modes of the gauge bosons for this
 class of models.
\end{references}
\end{document}